\documentclass[floatfix,lengthcheck,showpacs,amssymb,amsmath,amsfonts,twocolumn,nolinenumbers]{aastex63}
\usepackage{CJK}
\usepackage{lineno}
\usepackage[utf8]{inputenc}
\usepackage{color}
\usepackage{graphicx}
\usepackage{float}
\usepackage{subfigure}
\usepackage{amsmath}
\usepackage{afterpage}
\usepackage{acro}
\usepackage{multirow}
\usepackage{tikz}
\usetikzlibrary{arrows,shapes,trees,decorations.pathreplacing}
\usepackage{import}
\usepackage{svn-multi}
\usepackage{hyperref}

\DeclareAcronym{GR}{
	short = GR,
	long  = general relativity
	}
	
\DeclareAcronym{BH}{
	short = BH ,
	long  = black hole
}
\DeclareAcronym{SGWB}{
	short = SGWB ,
	long  = stochastic gravitational-wave background
}
\DeclareAcronym{BBH}{
	short = BBH ,
	long  = binary black hole
}
\DeclareAcronym{BNS}{
	short = BNS ,
	long  = binary neutron star
}
\DeclareAcronym{NSBH}{
	short = NSBH ,
	long  = neutron star black hole
}

\DeclareAcronym{GW}{
	short = GW ,
	long  = gravitational wave
}

\DeclareAcronym{CBC}{
	short = CBC,
	long  = compact binary coalescence
}

\DeclareAcronym{FRB}{
	short = {FRB},
	long  = fast radio burst
}

\DeclareAcronym{SNR}{
	short = SNR,
	long  = signal-to-noise ratio
}

\usepackage[capitalise]{cleveref}
\crefname{figure}{Fig.}{Figs.}
\Crefname{figure}{Fig.}{Figs.}

\def\be{\begin{equation}}
\def\ee{\end{equation}}
\def\({\left(}
\def\){\right)}
\def\[{\left[}
\def\]{\right]}

\begin{document}
\begin{CJK*}{UTF8}{gbsn}
\title{Search for Coincident Gravitational-wave and Fast Radio Burst Events from 4-OGC and the First CHIME/FRB Catalog}
\correspondingauthor{Yi-Fan Wang (王一帆)}
\email{yifan.wang@aei.mpg.de}

\author[0000-0002-2928-2916]{Yi-Fan Wang (王一帆)}
\author[0000-0002-1850-4587]{Alexander H. Nitz}
\affiliation{Max-Planck-Institut f{\"u}r Gravitationsphysik (Albert-Einstein-Institut), D-30167 Hannover, Germany}
\affiliation{Leibniz Universit{\"a}t Hannover, D-30167 Hannover, Germany}

\begin{abstract}
Advanced LIGO and Virgo have reported 90 confident gravitational-wave (GW) observations from compact-binary coalescences from their three observation runs. 
In addition, numerous subthreshold gravitational-wave candidates have been identified.
Binary neutron star (BNS) mergers can produce gravitational waves and short-gamma ray bursts, as confirmed by GW170817/GRB\,170817A. 
There may be electromagnetic counterparts recorded in archival observations associated with subthreshold gravitational-wave candidates. 
The CHIME/FRB Collaboration has reported the first large sample of fast radio bursts (FRBs), millisecond radio transients detected up to cosmological distances; a fraction of these may be associated with BNS mergers.
This work searches for coincident gravitational waves and FRBs from BNS mergers using candidates from the fourth-Open Gravitational-wave Catalog (4-OGC) and the first CHIME/FRB catalog. 
We use a ranking statistic for GW/FRB association that combines the gravitational-wave detection statistic with the odds of temporal and spatial association.
We analyze gravitational-wave candidates and non-repeating FRBs from 2019 April 1 to 2019 July 1, when both the Advanced LIGO/Virgo gravitational-wave detectors and the CHIME radio telescope were observing.
The most significant coincident candidate has a false alarm rate of 0.29 per observation time, which is consistent with a null observation.
The null results imply, at most, $\mathcal{O}(0.01)\%$ - $\mathcal{O}(1)\%$ of FRBs are produced promptly from the BNS mergers.

\end{abstract}

\acresetall

\section{Introduction} \label{sec:intro}

Gravitational-wave (GW) observations have opened up a new opportunity for multimessenger astronomy.
On 2017 August 17, the Advanced LIGO \citep{TheLIGOScientific:2014jea} and Virgo \citep{TheVirgo:2014hva} observatories detected a gravitational wave consistent with a binary neutron star (BNS) merger \citep[GW170817;][]{TheLIGOScientific:2017qsa,Monitor:2017mdv,GBM:2017lvd}.
An associated gamma-ray burst was independently detected by the Fermi Gamma-ray Burst Monitor \citep{2009ApJ...702..791M}. Subsequent observations found a broad spectrum of electromagnetic emissions, from X-ray,  gamma-ray to optical and radio \citep{GBM:2017lvd}.
The detection has confirmed short-gamma-ray bursts originate from BNS mergers \citep{1986ApJ...308L..47G,1986ApJ...308L..43P,cite-key,1992ApJ...395L..83N}, and has several implications such as measurements of neutron star properties \citep{LIGOScientific:2018cki,Capano:2019eae,Dietrich:2020efo,Landry:2020vaw} and inference of the Hubble constant
\citep{LIGOScientific:2017adf,LIGOScientific:2018gmd,Guidorzi:2017ogy,Hotokezaka:2018dfi}.

A potential electromagnetic counterpart to gravitational waves is a fast radio burst (FRB); a millisecond-duration radio transient found up to cosmological distances.
The first FRB event was identified in 2007 \citep{2007Sci...318..777L}. Later, FRB\,121102 \citep{2016Natur.531..202S} showed evidence of repeating FRBs, suggesting a classification of repeaters and apparent non-repeaters.
The possible generation mechanisms for FRBs remain a topic of discussion (for a review, see, e.g., \cite{Zhang:2020qgp}).
The detection of a galactic repeating fast radio burst FRB\,200428 from a magnetar \citep{CHIMEFRB:2020abu,Bochenek:2020zxn} associated with hard X-rays and soft gamma-rays \citep{Insight-HXMTTeam:2020dmu,Mereghetti:2020unm,Ridnaia:2020gcv,Tavani:2020adq} confirms at least a portion of FRBs originate from magnetars.
Nevertheless, some FRBs appear not to repeat, and the origin of non-repeaters remains under active investigation.
In addition to the magnetar flare activities \citep{Popov:2007uv,Metzger_2017,10.1093/mnras/stz700,Margalit_2020}, possible mechanisms to produce FRBs include the prompt emission from BNS mergers \citep{2013PASJ...65L..12T, Wang:2016dgs,Sridhar:2020uez}, collisions of neutron stars with asteroids or comets \citep{Geng:2015vza}, giant pulses from pulsars \citep{2012MNRAS.425L..71K}, or more exotic scenarios such as cosmic strings \citep{PhysRevLett.101.141301}.
Of particular interest to this work is the prompt emission of FRBs during the BNS merger because of the possible gravitational-wave counterpart.

Since entering the advanced detector era, LIGO and Virgo have completed three observation runs. The LVK collaboration has reported 90 gravitational-wave observations from compact-binary coalescences up to the third  Gravitational-Wave Transient Catalog (GWTC-3) \citep{LIGOScientific:2021djp}. 
Independent groups have searched the public data \citep{Vallisneri:2014vxa, Abbott:2019ebz} and reported additional detections \citep{Nitz:2021zwj, Olsen:2022pin}. \cite{Nitz:2021zwj} has reported the fourth Open Gravitational-wave Catalog (4-OGC) based on the \texttt{PyCBC} pipeline \citep{pycbc-github}. In this work, we analyze the public set of subthreshold candidates from the 4-OGC catalog \citep{Nitz:2021zwj}.

The Canadian Hydrogen Intensity Mapping Experiment~\citep[CHIME;][]{CHIMEFRB:2018mlh} has released the first CHIME/FRB catalog~\citep{CHIMEFRB:2021srp}, reporting 536 events within 1 yr of observation.
For 3 months, from April 1 to July 1 in 2019, the gravitational-wave detectors Advanced LIGO/Virgo and CHIME were both observing, establishing the opportunity to search for coincident signals.
The data obtained from a single uniform survey also enables straightforward background noise estimation for the GW/FRB association.
To examine the hypothesized BNS merger scenario for FRBs, we identify gravitational wave triggers within a +/- 100 s around each FRB in the CHIME catalog.
We compute a statistic for each pair of temporally coincident events based on the gravitational-wave detection statistic and the Bayes' factor measuring the sky localization association. To estimate the false alarm rate of our search, we empirically measure the background noise by artificially sliding the time of CHIME events with respect to LIGO/Virgo by a temporal stride of $\sim$1 day such that we assume any spatial and temporal association is due to false coincidences.

We analyze the available candidates from the 4-OGC and CHIME/FRB catalog that pass our selection criteria; the most significant associated GW/FRB candidate has a false alarm rate of 0.29 per observation time, $\sim$2.4 months (the duty cycle that at least two GW detectors were observing is 81\% from April 1 to July 1 in 2019 and we do not consider single detector GW triggers \citep{Nitz:2020naa}). 
{We adopt criteria from the GW community that a false alarm rate smaller than 1 per 100 yr is considered as a significant detection, which is used by, e.g., \cite{PhysRevX.6.041015}, } and thus conclude a null result in our GW/FRB search. 
Our result is consistent with the recent null detection from \cite{2022arXiv220312038T}.
We estimate the fraction of FRBs that are associated with BNS mergers is at most $\mathcal{O}(0.01)\%$ - $\mathcal{O}(1)\%$.

\section{Methods for coincident GW/FRB search}\label{sec:con}

This section briefly reviews the Bayesian inference of gravitational wave events, the selection criteria for GW/FRB candidates, and the statistic to rank the GW/FRB association.

\subsection{Bayesian inference for gravitational wave}
Given gravitational wave time series data ${d(t)}$ which is a sum of the detector noise $n(t)$ and a gravitational wave signal $h(t,\vec{\theta})$ with characterizing parameters $\vec{\theta}$, Bayes' theorem states that
\be\label{eq:pe}
P(\vec{\theta}|d,H) = \frac{ P(d|\vec{\theta},H) P(\vec{\theta}|H)} {P(d|H)},
\ee
where $P(\vec{\theta}|d,H)$ is the posterior probability distribution for parameters $\vec{\theta}$, $P(\vec{\theta}|H)$ is the prior distribution for $\vec{\theta}$, $P(d | \vec{\theta},H)$ is the likelihood to obtain the data given a set of model parameters, and $P(d|H)$ is a normalization factor called the evidence, {which is an integral of the prior-weighted likelihood marginalized over all the model parameters $\vec\theta$}
\be
P(d|H) = \int P(d|\vec\theta,H)P(\vec\theta|H)\mathrm{d}\vec\theta.
\ee
$H$ is the underlying hypothesis characterizing the signal detection.
The Bayes odds ratio, or the ratio of the evidence of two hypotheses, is 
\begin{equation}
\mathcal{B}^1_2 = \frac{P(d|H_1)}{P(d|H_2)},
\end{equation} 
which quantitatively measures the degree to which hypothesis the data favors.

For Gaussian and stationary noise in gravitational wave detectors, the likelihood function is
\be \label{eq:likelihood}
P(d|\vec{\theta}, H) \propto \exp\[	-\frac{1}{2}\sum_{i} \langle d_i-h_i(\vec{\theta})|d_i-h_i(\vec{\theta})\rangle \],
\ee
where {$h_i(\vec{\theta})$} is the waveform template of gravitational wave given by model $H$ {and parameters $\vec\theta$}, and $i$ represents the i-th gravitational wave detector.
The inner product $\langle a|b\rangle$ is defined to be
\be
\langle a|b\rangle = 4 \mathfrak{R} \int \frac{{a}(f){b}^*(f)}{S_h(f)} df,
\ee
where $S_h(f)$ is the one-side noise power spectral density of the gravitational wave detector as a function of frequency $f$.

\subsection{Candidate selection}
To search for possible GW/FRB associations, we first select the FRBs from the CHIME/FRB catalog occurring from 2019 April 1 to 2019 July 1; this time range overlaps with the first half of the Advanced LIGO/Virgo third observation run.
Since the prompt emissions of FRBs associated with a BNS merger are not expected to repeat \citep{2013PASJ...65L..12T, Wang:2016dgs,Sridhar:2020uez}, we only select the apparent non-repeaters.
One hundred fifty-one FRB events remain from the selection.

The targeted search for gravitational waves from compact-binary coalescence relies on accurately modeling the expected signals to enable their use as templates~\citep{Harry:2009ea}.
The strain data from gravitational-wave detectors are match filtered against a bank of templates, and then a detection statistic is assigned for each candidate~\citep{Allen:2005fk}.
We use subthreshold gravitational-wave candidates recorded in the public release of the 4-OGC catalog \citep{Nitz:2021zwj}. The binary component mass and spin parameters from the best-ranked template are given for each candidate.

We select the gravitational-wave candidates, again occurring from 2019 April 1 to 2019 July 1, whose component mass ranges from one to two solar masses.
We consider gravitational-wave candidates passing these criteria to potentially arise from BNS mergers.
{The search template bank for BNSs in 4-OGC targets the magnitude of dimensionless spin in [-0.05, 0.05] aligned with orbital angular momentum \citep{Nitz:2021zwj}, because BNSs are not expected to have large spins during merger as suggested by the known gravitational-wave sources \citep{LIGOScientific:2018hze, Abbott:2020uma} and galactic BNSs \citep{PhysRevD.98.043002}.
The search would be sensitive to higher spin sources though because of degeneracy between the effective spin and mass ratio for the GW waveform.
Extending the search template bank to cover spin up to $0.5$ would require an order of magnitude more templates, and trigger the same scale of more background noise.
The search approximately loses $\sim 10\%$ sensitive volume for every order of magnitude in background noise \citep{Brown:2012qf}.
Overall, } around $3\times10^5$ subthreshold GW candidates satisfy the selection criteria.

\subsection{Ranking statistic for GW/FRB association}

We rank each GW/FRB pair by combining the GW candidate's ranking statistic with the odds that 
both observations occur at nearly the same time and from a common sky location, following the same spirit of \cite{Nitz:2019bxt} and \cite{2018ApJ...860....6A}. This can be expressed as the sum of the GW detection statistic $\lambda_\mathrm{gw}$ \citep{Nitz:2017svb,Davies:2020tsx} and $\ln \mathcal{B}^\mathrm{fixsky}_\mathrm{relaxsky}$:
\begin{equation}
\lambda_\mathrm{gw+frb} = \lambda_\mathrm{gw} + \ln \mathcal{B}^\mathrm{fixsky}_\mathrm{relaxsky}.
\end{equation}
The expression of $\lambda_\mathrm{gw}$ for \texttt{PyCBC} is in, e.g., Eq. (16) of \cite{Davies:2020tsx}, which represents the natural logarithm of the ratio between signal rate density and noise rate density for a gravitational wave trigger.
The ranking statistic itself has a clear physical meaning that combines a detection statistic for gravitational wave triggers and an improvement of Bayesian evidence if the gravitational wave trigger is indeed from the direction informed by an FRB. We assume that each FRB observation does not arise from noise and so do not include any additional factors for its likelihood.

To obtain  $\ln \mathcal{B}^\mathrm{fixsky}_\mathrm{relaxsky}$, we first look for temporally coincident pairs by selecting gravitational wave triggers occurring within a +/- 100 seconds window with respect to the epoch of an FRB event using $\texttt{mjd\_inf}$ given by the CHIME/FRB catalog \citep{CHIMEFRB:2021srp}.
This time window accounts for the time delay of an FRB due to ionized gas in interstellar and intergalactic media, which is $\mathcal{O}(10)s$ \citep{James:2019xca}.
The sky localization of FRBs is much more precise than for each gravitational-wave candidate; the uncertainty for right ascension and declination in the CHIME/FRB catalog is at the sub-degree level.  In contrast, the sky localization uncertainty for gravitational waves is tens to hundreds of square degrees. In this work, we use only the two-dimensional sky position information and do not explicitly account for the GW/FRB distance consistency. 
We define two competing hypotheses using the gravitational-wave data.

\begin{quote}
$\mathcal{H}_1$: The sky location of the gravitational-wave observation is fixed to the right ascension and declination informed by the temporally associated FRB event;

$\mathcal{H}_2$:  The sky location of the gravitational-wave observation is considered unknown with a prior expectation that sources are isotropically distributed.
\end{quote}
The Bayesian evidence for the two hypotheses are $P(d|\mathcal{H}_1)$ and $P(d|\mathcal{H}_2)$, respectively, and the natural logarithm of the ratio of $P(d|\mathcal{H}_1)$ and $P(d|\mathcal{H}_2)$ is defined to be $\ln \mathcal{B}^\mathrm{fixsky}_\mathrm{relaxsky}$, which quantifies the preference for a gravitational wave trigger to originate from the same sky position of the associated FRB.

We use \texttt{TaylorF2} \citep{Sathyaprakash:1991mt,Droz:1999qx,Blanchet:2002av,Faye:2012we} to model the gravitational-wave signal $h(\vec{\theta})$ and use the dynamic nested sampler \texttt{Dynesty} \citep{speagle:2019} in \texttt{PyCBC Inference} \citep{Biwer:2018osg} to numerically compute the Bayesian evidence.
To simplify the calculation, we fix the component masses and spins of the gravitational-wave signal model to the values given by the best-ranked templates reported by 4-OGC.
We do not expect the sky location will be significantly biased by fixing the mass and spin given the decoupling of intrinsic and extrinsic parameters~\citep{PhysRevD.93.024013}.
Therefore, the variables to infer are the luminosity distance, inclination angle between the total angular momentum direction of the BNS and the line of sight toward the detectors, polarization angle, coalescence time, and phase, for the model $\mathcal{H}_1$.
The hypothesis $H_2$ adds two more variables, the right ascension, and declination.
The prior for luminosity distance is from a uniform distribution of comoving volume and is up to $450$ Mpc.
The priors for angular variables are isotropically distributed.

To determine the false alarm rate of our search as a function of our ranking statistic, we need the background of chance GW/FRB associations.
The estimation is achieved by artificially shifting the time of FRB events with respect to the gravitational wave triggers by a time much larger than 100 s (we choose $\sim$ 1 day as a stride). Since we have assumed that GW/FRB can only be associated within the +/- 100 s window, any coincident association after time sliding is due to a false coincidence. Using this method, we simulate $\sim$20 yr of background.

\section{Search results and implications}

\begin{figure}[htp]
    \centering
    \includegraphics[width=\columnwidth]{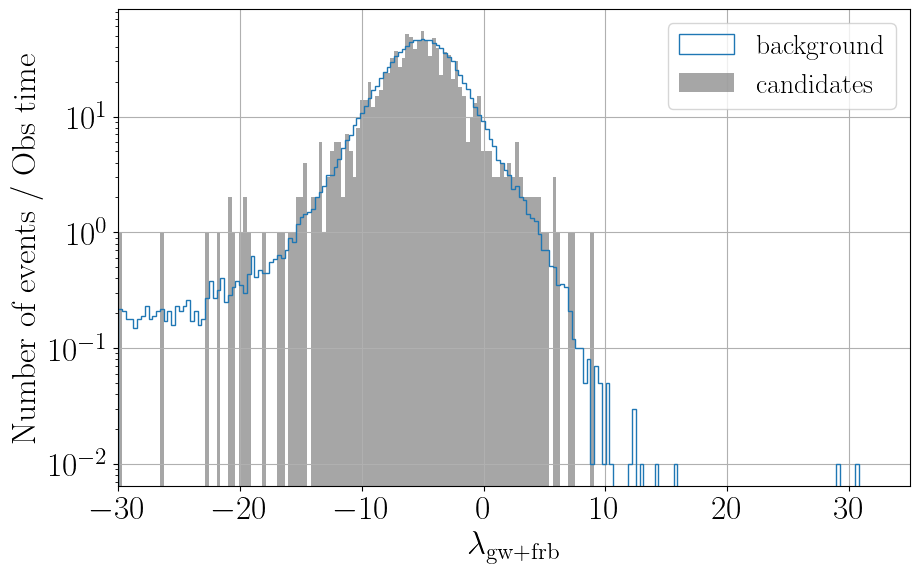}
    \caption{Histogram of the detection statistic values for our search candidates and background noise. The y-axis represents the number of events per observation time of 2.4 months, and the bin width is 0.3. The most significant trigger has a $\lambda_\mathrm{gw+frb}$ = 8.92 with a false alarm rate 0.29 per observation time.
    }
    \label{fig: searches}
\end{figure}

We compute $\lambda_\mathrm{gw+frb}$ for all associated pairs of GW/FRBs.
The search candidates and estimated background noise are shown in \cref{fig: searches}.
The gravitational-wave event name, the fast radio burst events name, the detection statistic, and the false alarm rate for the top three significant candidates are given in \cref{table}.
A complete list of candidates is provided in the associated data release~\citep{datarelease}.

\begin{table}[htp]
\begin{center}
\begin{tabular}{c|c|c|c|c}
GW Triggers & FRB Events  &$\ln \mathcal{B}^\mathrm{fixsky}_\mathrm{relaxsky}$ &$\lambda_\mathrm{gw+frb}$ &FAR/obs \\
\hline
 190701\_223118 & FRB\,20190701E &3.97 & 8.92&0.29	 \\ 
 190605\_021909 & FRB\,20190605C &3.64 &7.50&0.63 \\ 
 190411\_050213 & FRB\,20190411B& 3.27 & 6.95&0.93\\ 
\end{tabular}
\end{center}
\caption{The gravitational-wave trigger name, the fast radio burst events, the detection statistic and false alarm rate (FAR) per observation time, which is 2.4 months, for the top three significant candidates.}
\label{table}
\end{table}

\begin{figure}[htp]
    \centering
    \includegraphics[width=\columnwidth]{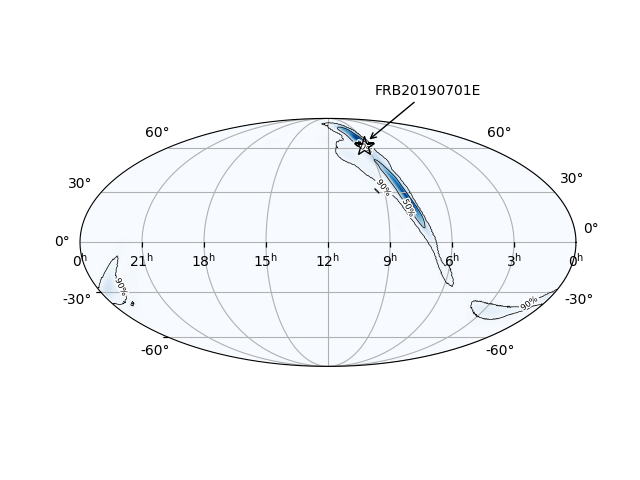}
    \vspace{-1cm}
    \caption{The posterior distribution of sky location for $190701\_223118$ from Bayesian inference given hypothesis $\mathcal{H}_2$, which assumes the right ascension and declination are free parameters to be inferred by gravitational-wave strain data. The $90\%$ and $50\%$ credible regions are shown. The marker represents the sky location of FRB\,20190701E measured by the CHIME/FRB collaboration.
    }
    \label{fig: sky}
\end{figure}

The most significant candidate is from the gravitational wave trigger $190701\_223118$ (the name is in the format of YYMMDD\_HHMMSS for UTC time) associated with FRB 20190701E which occurred 22 s later than the gravitational-wave candidate. The coincident candidate has $\textrm{ln}\mathcal{B}^\mathrm{fixsky}_\mathrm{relaxsky} \sim 3.97$.
\cref{fig: sky} shows the posterior of sky location for $190701\_223118$ inferred from the LIGO/Virgo data along with the sky location reported for FRB\,20190701E \citep{FRB20190701E}.
However, the associated false alarm rate is 0.29 per the (2.4 months) observation time.
In addition, if we further assume the interstellar and intergalactic medium model given by \cite{James:2019xca}, the dispersion measure of FRB\,20190701E \citep{FRB20190701E} corresponds to a luminosity distance of $\sim 3.6$ Gpc; in contrast the gravitational-wave posterior for luminosity distance gives $300^{+100}_{-120}$ Mpc for $190701\_223118$. We find no confident associations between GW and FRB candidates from the 4-OGC and CHIME/FRB catalogs.

Given the null search results, we constrain the rate of coincident GW/FRB observations.
{\cite{Callister:2016vtl} has estimated the physical FRB rate per comoving volume as (their Eq. 1)
\be
R_\mathrm{FRB}  = 3r_\mathrm{obs} / (\eta \Omega D^3 )
\ee
where $r_\mathrm{obs}$ is the observed FRB rate, $D$ is the comoving distance encompassing the observed FRB, $\eta$ is the detection efficiency in the range $[0,1]$ and $\Omega$ is the opening solid angle of FRB beam from BNS mergers.
Based on the observed FRB events and the beam area of the FAST radio telescope, \cite{Niu:2021xug} estimated the all-sky FRB rate to be $r_\mathrm{obs}= 10^5$ day$^{-1}$  sky$^{-1}$.
We take $D=5.4$ Gpc from the most distant FRB detected by \cite{Niu:2021xug}.
Thus the lowest event rate density for FRBs is $5.6\times10^4$  Gpc$^{-3}$ yr$^{-1}$ assuming the most optimal $100\%$ detection efficiency and $\Omega=4\pi$. 
Taking a more realistic assumption given by \cite{Callister:2016vtl} that takes account for a lower detection efficiency $\eta=50\%$ and an opening angle $30^\circ$ ($\Omega\simeq0.8$), the FRB event rate density increases to $1.7\times10^6$ Gpc$^{-3}$ yr$^{-1}$.}

The BNS merger rate from the 4-OGC catalog is estimated to be $200^{+309}_{-148}$ Gpc$^{-3}$ yr$^{-1}$  \citep[$90\%$ credible intervals;][]{Nitz:2021zwj} if we consider GW170817 \citep{TheLIGOScientific:2017qsa} and GW190425 \citep{Abbott:2020uma} to be representative members of a standard BNS population.
We {thus take $509$ Gpc$^{-3}$ yr$^{-1}$} to be the upper limit of event rate density of BNS mergers that are associated with FRBs. 
Taking the lowest FRB event rate density indicates at most $\mathcal{O}(1\%)$ of FRBs can originate from prompt emission of BNS mergers.
A more realistic estimation for FRB event rate suggests BNS mergers can only take account for $\mathcal{O}(0.01\%)$ of FRBs.

\section{Discussion and Conclusion}

We have searched for temporally and spatially coincident gravitational-wave and FRB candidates from the public 4-OGC and CHIME/FRB catalogs.
Based on a ranking statistic accounting for the gravitational-wave trigger significance and the odds of sky position association, the most significant coincident pair has a false alarm rate of 0.29 per observation time. 
Thus no confident detections are made.
We estimate up to  $\mathcal{O}(0.01)\%$ - $\mathcal{O}(1)\%$ of FRBs can be accounted for by BNS mergers through an order of magnitude comparison of the observed BNS merger rate from gravitational-wave observations and estimates of the FRB event rate.
The most optimal fraction $\mathcal{O}(1)\%$ implies $\sim\mathcal{O}(1)$ FRBs in our search are expected to be accompanied by a gravitational-wave signal; however, it may be beyond the horizon distance of the current gravitational-wave observatories.

Improved understanding and modeling of the joint properties of a potential GW/FRB observation may be able to further increase the sensitivity of future analyses. 
For example, the time delay between the associated GW and FRB was chosen to be uniform in a +/- 100 s window. Future work may enable a more realistic time delay distribution that can be accounted for directly in the ranking statistic. 
We did not consider the detection probability of each FRB in the statistic, effectively setting all events on the same footing and assuming that none arise from noise.
In addition, if only a small portion of FRBs arise from prompt BNS mergers, there may be features in the observed emission that enable distinguishing FRB mechanisms; we currently limit to only excluding repeating sources. 
Future work may also be able to take into account the FRB dispersion measure into a combined statistic. 

Our detection statistic can also be straightforwardly applied to search for coincident binary black hole mergers and FRBs, given models that predict charged black hole mergers may produce FRBs \citep{Zhang:2016rli,Liu:2016olx}.
Future work can also search gravitational waves associated with repeating FRBs powered by a neutron star formed after the BNS merger, as predicted by \cite{Wang:2020aut} and \cite{Zhang:2020cfp}.
Multimessenger searches could also include neutrinos from the IceCube catalogs \citep{IceCube:2021xar} or gamma-ray burst counterparts \citep{Nitz:2019bxt, LIGOScientific:2021iyk, LIGOScientific:2020lst}.

The fourth observation run of Advanced LIGO/Virgo/KAGRA is scheduled to start on 2022 December \citep{Aasi:2013wya}; the BNS horizon distance is expected to expand to 160 - 190 Mpc for LIGO, compared with the horizon distance of 120 - 140 Mpc in the third observation run \citep{LIGOScientific:2021djp}. The improved detectors have the potential to observe GW170817-like multimessenger observations and provide tighter constraints on the BNS merger/FRB association event rate or even make the first detection for gravitational wave and FRB multimessenger astronomy.

A complete list of search candidates and scripts associated with this work are released in \cite{datarelease}.

\acknowledgements
We acknowledge the Max Planck Gesellschaft and the Atlas cluster computing team at AEI Hannover for support.  This research has made use of data, software and/or web tools obtained from the Gravitational Wave Open Science Center (https://www.gw-openscience.org), a service of LIGO Laboratory, the LIGO Scientific Collaboration and the Virgo Collaboration. LIGO is funded by the U.S. National Science Foundation. Virgo is funded by the French Centre National de Recherche Scientifique (CNRS), the Italian Istituto Nazionale della Fisica Nucleare (INFN) and the Dutch Nikhef, with contributions by Polish and Hungarian institutes.

\bibliography{reference}

\begin{thebibliography}{}
\expandafter\ifx\csname natexlab\endcsname\relax\def\natexlab#1{#1}\fi
\providecommand{\url}[1]{\href{#1}{#1}}
\providecommand{\dodoi}[1]{doi:~\href{http://doi.org/#1}{\nolinkurl{#1}}}
\providecommand{\doeprint}[1]{\href{http://ascl.net/#1}{\nolinkurl{http://ascl.net/#1}}}
\providecommand{\doarXiv}[1]{\href{https://arxiv.org/abs/#1}{\nolinkurl{https://arxiv.org/abs/#1}}}

\bibitem[{Aasi {et~al.}(2015)}]{TheLIGOScientific:2014jea}
Aasi, J., {et~al.} 2015, Class. Quantum Grav., 32, 074001,
  \dodoi{10.1088/0264-9381/32/7/074001}

\bibitem[{Abbasi {et~al.}(2021)}]{IceCube:2021xar}
Abbasi, R., {et~al.} 2021, \dodoi{10.21234/CPKQ-K003}

\bibitem[{Abbott {et~al.}(2017{\natexlab{a}})}]{TheLIGOScientific:2017qsa}
Abbott, B., {et~al.} 2017{\natexlab{a}}, Phys. Rev. Lett., 119, 161101,
  \dodoi{10.1103/PhysRevLett.119.161101}

\bibitem[{Abbott {et~al.}(2017{\natexlab{b}})}]{Monitor:2017mdv}
---. 2017{\natexlab{b}}, Astrophys. J. Lett., 848, L13,
  \dodoi{10.3847/2041-8213/aa920c}

\bibitem[{Abbott {et~al.}(2020)}]{Abbott:2020uma}
---. 2020, Astrophys. J. Lett., 892, L3, \dodoi{10.3847/2041-8213/ab75f5}

\bibitem[{Abbott {et~al.}(2016{\natexlab{a}})Abbott, Abbott, Abbott, Abernathy,
  Acernese, Ackley, Adams, Adams, Addesso, Adhikari, Adya, Affeldt, Agathos,
  Agatsuma, Aggarwal, Aguiar, Aiello, Ain, Ajith, Allen, Allocca, Altin,
  Anderson, Anderson, Arai, Araya, Arceneaux, Areeda, Arnaud, Arun, Ascenzi,
  Ashton, Ast, Aston, Astone, Aufmuth, Aulbert, Babak, Bacon, Bader, Baker,
  Baldaccini, Ballardin, Ballmer, Barayoga, Barclay, Barish, Barker, Barone,
  Barr, Barsotti, Barsuglia, Barta, Bartlett, Bartos, Bassiri, Basti, Batch,
  Baune, Bavigadda, Bazzan, Bejger, Bell, Berger, Bergmann, Berry, Bersanetti,
  Bertolini, Betzwieser, Bhagwat, Bhandare, Bilenko, Billingsley, Birch,
  Birney, Birnholtz, Biscans, Bisht, Bitossi, Biwer, Bizouard, Blackburn,
  Blair, Blair, Blair, Bloemen, Bock, Boer, Bogaert, Bogan, Bohe, Bond, Bondu,
  Bonnand, Boom, Bork, Boschi, Bose, Bouffanais, Bozzi, Bradaschia, Brady,
  Braginsky, Branchesi, Brau, Briant, Brillet, Brinkmann, Brisson, Brockill,
  Broida, Brooks, Brown, Brown, Brown, Brunett, Buchanan, Buikema, Bulik,
  Bulten, Buonanno, Buskulic, Buy, Byer, Cabero, Cadonati, Cagnoli, Cahillane,
  Calder\'on~Bustillo, Callister, Calloni, Camp, Cannon, Cao, Capano, Capocasa,
  Carbognani, Caride, Casanueva~Diaz, Casentini, Caudill, Cavagli\`a, Cavalier,
  Cavalieri, Cella, Cepeda, Cerboni~Baiardi, Cerretani, Cesarini, Chamberlin,
  Chan, Chao, Charlton, Chassande-Mottin, Cheeseboro, Chen, Chen, Cheng,
  Chincarini, Chiummo, Cho, Cho, Chow, Christensen, Chu, Chua, Chung, Ciani,
  Clara, Clark, Cleva, Coccia, Cohadon, Colla, Collette, Cominsky, Constancio,
  Conte, Conti, Cook, Corbitt, Cornish, Corsi, Cortese, Costa, Coughlin,
  Coughlin, Coulon, Countryman, Couvares, Cowan, Coward, Cowart, Coyne, Coyne,
  Craig, Creighton, Cripe, Crowder, Cumming, Cunningham, Cuoco, Dal~Canton,
  Danilishin, D'Antonio, Danzmann, Darman, Dasgupta, Da~Silva~Costa, Dattilo,
  Dave, Davier, Davies, Daw, Day, De, DeBra, Debreczeni, Degallaix,
  De~Laurentis, Del\'eglise, Del~Pozzo, Denker, Dent, Dergachev, De~Rosa,
  DeRosa, DeSalvo, Devine, Dhurandhar, D\'{\i}az, Di~Fiore, Di~Giovanni,
  Di~Girolamo, Di~Lieto, Di~Pace, Di~Palma, Di~Virgilio, Dolique, Donovan,
  Dooley, Doravari, Douglas, Downes, Drago, Drever, Driggers, Ducrot, Dwyer,
  Edo, Edwards, Effler, Eggenstein, Ehrens, Eichholz, Eikenberry, Engels,
  Essick, Etzel, Evans, Evans, Everett, Factourovich, Fafone, Fair, Fairhurst,
  Fan, Fang, Farinon, Farr, Farr, Favata, Fays, Fehrmann, Fejer, Fenyvesi,
  Ferrante, Ferreira, Ferrini, Fidecaro, Fiori, Fiorucci, Fisher, Flaminio,
  Fletcher, Fong, Fournier, Frasca, Frasconi, Frei, Freise, Frey, Frey,
  Fritschel, Frolov, Fulda, Fyffe, Gabbard, Gaebel, Gair, Gammaitoni, Gaonkar,
  Garufi, Gaur, Gehrels, Gemme, Geng, Genin, Gennai, George, Gergely, Germain,
  Ghosh, Ghosh, Ghosh, Giaime, Giardina, Giazotto, Gill, Glaefke, Goetz, Goetz,
  Gondan, Gonz\'alez, Gonzalez~Castro, Gopakumar, Gordon, Gorodetsky, Gossan,
  Gosselin, Gouaty, Grado, Graef, Graff, Granata, Grant, Gras, Gray, Greco,
  Green, Groot, Grote, Grunewald, Guidi, Guo, Gupta, Gupta, Gushwa, Gustafson,
  Gustafson, Hacker, Hall, Hall, Hamilton, Hammond, Haney, Hanke, Hanks, Hanna,
  Hannam, Hanson, Hardwick, Harms, Harry, Harry, Hart, Hartman, Haster,
  Haughian, Healy, Heidmann, Heintze, Heitmann, Hello, Hemming, Hendry, Heng,
  Hennig, Henry, Heptonstall, Heurs, Hild, Hoak, Hofman, Holt, Holz, Hopkins,
  Hough, Houston, Howell, Hu, Huang, Huerta, Huet, Hughey, Husa, Huttner,
  Huynh-Dinh, Indik, Ingram, Inta, Isa, Isac, Isi, Isogai, Iyer, Izumi,
  Jacqmin, Jang, Jani, Jaranowski, Jawahar, Jian, Jim\'enez-Forteza, Johnson,
  Johnson-McDaniel, Jones, Jones, Jonker, Ju, K, Kalaghatgi, Kalogera,
  Kandhasamy, Kang, Kanner, Kapadia, Karki, Karvinen, Kasprzack, Katsavounidis,
  Katzman, Kaufer, Kaur, Kawabe, K\'ef\'elian, Kehl, Keitel, Kelley, Kells,
  Kennedy, Key, Khalili, Khan, Khan, Khan, Khazanov, Kijbunchoo, Kim, Kim, Kim,
  Kim, Kim, Kim, Kim, Kimbrell, King, King, Kissel, Klein, Kleybolte, Klimenko,
  Koehlenbeck, Koley, Kondrashov, Kontos, Korobko, Korth, Kowalska, Kozak,
  Kringel, Krishnan, Kr\'olak, Krueger, Kuehn, Kumar, Kumar, Kuo, Kutynia,
  Lackey, Landry, Lange, Lantz, Lasky, Laxen, Lazzarini, Lazzaro, Leaci,
  Leavey, Lebigot, Lee, Lee, Lee, Lee, Lenon, Leonardi, Leong, Leroy, Letendre,
  Levin, Lewis, Li, Libson, Littenberg, Lockerbie, Lombardi, London, Lord,
  Lorenzini, Loriette, Lormand, Losurdo, Lough, Lousto, L\"uck, Lundgren,
  Lynch, Ma, Machenschalk, MacInnis, Macleod, Maga\~na Sandoval, Maga\~na
  Zertuche, Magee, Majorana, Maksimovic, Malvezzi, Man, Mandel, Mandic,
  Mangano, Mansell, Manske, Mantovani, Marchesoni, Marion, M\'arka, M\'arka,
  Markosyan, Maros, Martelli, Martellini, Martin, Martynov, Marx, Mason,
  Masserot, Massinger, Masso-Reid, Mastrogiovanni, Matichard, Matone,
  Mavalvala, Mazumder, McCarthy, McClelland, McCormick, McGuire, McIntyre,
  McIver, McManus, McRae, McWilliams, Meacher, Meadors, Meidam, Melatos,
  Mendell, Mercer, Merilh, Merzougui, Meshkov, Messenger, Messick, Metzdorff,
  Meyers, Mezzani, Miao, Michel, Middleton, Mikhailov, Milano, Miller, Miller,
  Miller, Miller, Millhouse, Minenkov, Ming, Mirshekari, Mishra, Mitra,
  Mitrofanov, Mitselmakher, Mittleman, Moggi, Mohan, Mohapatra, Montani, Moore,
  Moore, Moraru, Moreno, Morriss, Mossavi, Mours, Mow-Lowry, Mueller, Muir,
  Mukherjee, Mukherjee, Mukherjee, Mukund, Mullavey, Munch, Murphy, Murray,
  Mytidis, Nardecchia, Naticchioni, Nayak, Nedkova, Nelemans, Nelson, Neri,
  Neunzert, Newton, Nguyen, Nielsen, Nissanke, Nitz, Nocera, Nolting,
  Normandin, Nuttall, Oberling, Ochsner, O'Dell, Oelker, Ogin, Oh, Oh, Ohme,
  Oliver, Oppermann, Oram, O'Reilly, O'Shaughnessy, Ottaway, Overmier, Owen,
  Pai, Pai, Palamos, Palashov, Palomba, Pal-Singh, Pan, Pan, Pankow, Pannarale,
  Pant, Paoletti, Paoli, Papa, Paris, Parker, Pascucci, Pasqualetti,
  Passaquieti, Passuello, Patricelli, Patrick, Pearlstone, Pedraza, Pedurand,
  Pekowsky, Pele, Penn, Perreca, Perri, Pfeiffer, Phelps, Piccinni, Pichot,
  Piergiovanni, Pierro, Pillant, Pinard, Pinto, Pitkin, Poe, Poggiani,
  Popolizio, Porter, Post, Powell, Prasad, Predoi, Prestegard, Price,
  Prijatelj, Principe, Privitera, Prix, Prodi, Prokhorov, Puncken, Punturo,
  Puppo, P\"urrer, Qi, Qin, Qiu, Quetschke, Quintero, Quitzow-James, Raab,
  Rabeling, Radkins, Raffai, Raja, Rajan, Rakhmanov, Rapagnani, Raymond,
  Razzano, Re, Read, Reed, Regimbau, Rei, Reid, Reitze, Rew, Reyes, Ricci,
  Riles, Rizzo, Robertson, Robie, Robinet, Rocchi, Rolland, Rollins, Roma,
  Romano, Romano, Romanov, Romie, Rosi\ifmmode~\acute{n}\else \'{n}\fi{}ska,
  Rowan, R\"udiger, Ruggi, Ryan, Sachdev, Sadecki, Sadeghian, Sakellariadou,
  Salconi, Saleem, Salemi, Samajdar, Sammut, Sanchez, Sandberg, Sandeen,
  Sanders, Sassolas, Sathyaprakash, Saulson, Sauter, Savage, Sawadsky, Schale,
  Schilling, Schmidt, Schmidt, Schnabel, Schofield, Sch\"onbeck, Schreiber,
  Schuette, Schutz, Scott, Scott, Sellers, Sengupta, Sentenac, Sequino,
  Sergeev, Setyawati, Shaddock, Shaffer, Shahriar, Shaltev, Shapiro, Shawhan,
  Sheperd, Shoemaker, Shoemaker, Siellez, Siemens, Sieniawska, Sigg, Silva,
  Singer, Singer, Singh, Singh, Singhal, Sintes, Slagmolen, Smith, Smith,
  Smith, Son, Sorazu, Sorrentino, Souradeep, Srivastava, Staley, Steinke,
  Steinlechner, Steinlechner, Steinmeyer, Stephens, Stevenson, Stone, Strain,
  Straniero, Stratta, Strauss, Strigin, Sturani, Stuver, Summerscales, Sun,
  Sunil, Sutton, Swinkels, Szczepa\ifmmode~\acute{n}\else \'{n}\fi{}czyk,
  Tacca, Talukder, Tanner, T\'apai, Tarabrin, Taracchini, Taylor, Theeg,
  Thirugnanasambandam, Thomas, Thomas, Thomas, Thorne, Thrane, Tiwari, Tiwari,
  Tokmakov, Toland, Tomlinson, Tonelli, Tornasi, Torres, Torrie, T\"oyr\"a,
  Travasso, Traylor, Trifir\`o, Tringali, Trozzo, Tse, Turconi, Tuyenbayev,
  Ugolini, Unnikrishnan, Urban, Usman, Vahlbruch, Vajente, Valdes, Vallisneri,
  van Bakel, van Beuzekom, van~den Brand, Van Den~Broeck, Vander-Hyde, van~der
  Schaaf, van Heijningen, van Veggel, Vardaro, Vass, Vas\'uth, Vaulin, Vecchio,
  Vedovato, Veitch, Veitch, Venkateswara, Verkindt, Vetrano, Vicer\'e,
  Vinciguerra, Vine, Vinet, Vitale, Vo, Vocca, Vorvick, Voss, Vousden,
  Vyatchanin, Wade, Wade, Wade, Walker, Wallace, Walsh, Wang, Wang, Wang, Wang,
  Wang, Ward, Warner, Was, Weaver, Wei, Weinert, Weinstein, Weiss, Wen,
  We\ss{}els, Westphal, Wette, Whelan, Whitcomb, Whiting, Williams, Williamson,
  Willis, Willke, Wimmer, Winkler, Wipf, Wittel, Woan, Woehler, Worden, Wright,
  Wu, Wu, Yablon, Yam, Yamamoto, Yancey, Yu, Yvert, Zadro\ifmmode~\dot{z}\else
  \.{z}\fi{}ny, Zangrando, Zanolin, Zendri, Zevin, Zhang, Zhang, Zhang, Zhao,
  Zhou, Zhou, Zhu, Zucker, Zuraw, \& Zweizig}]{PhysRevX.6.041015}
Abbott, B.~P., Abbott, R., Abbott, T.~D., {et~al.} 2016{\natexlab{a}}, Phys.
  Rev. X, 6, 041015, \dodoi{10.1103/PhysRevX.6.041015}

\bibitem[{Abbott {et~al.}(2016{\natexlab{b}})}]{Aasi:2013wya}
Abbott, B.~P., {et~al.} 2016{\natexlab{b}}, Living Rev. Relat., 19, 1,
  \dodoi{10.1007/lrr-2016-1}

\bibitem[{Abbott {et~al.}(2017{\natexlab{c}})}]{GBM:2017lvd}
---. 2017{\natexlab{c}}, Astrophys. J., 848, L12,
  \dodoi{10.3847/2041-8213/aa91c9}

\bibitem[{Abbott {et~al.}(2017{\natexlab{d}})}]{LIGOScientific:2017adf}
---. 2017{\natexlab{d}}, Nature, 551, 85, \dodoi{10.1038/nature24471}

\bibitem[{Abbott {et~al.}(2018)}]{LIGOScientific:2018cki}
---. 2018, Phys. Rev. Lett., 121, 161101,
  \dodoi{10.1103/PhysRevLett.121.161101}

\bibitem[{Abbott {et~al.}(2019{\natexlab{a}})}]{LIGOScientific:2018hze}
---. 2019{\natexlab{a}}, Phys. Rev. X, 9, 011001,
  \dodoi{10.1103/PhysRevX.9.011001}

\bibitem[{Abbott {et~al.}(2019{\natexlab{b}})}]{Abbott:2019ebz}
Abbott, R., {et~al.} 2019{\natexlab{b}}.
\newblock \doarXiv{1912.11716}

\bibitem[{Abbott {et~al.}(2021{\natexlab{a}})}]{LIGOScientific:2021djp}
---. 2021{\natexlab{a}}.
\newblock \doarXiv{2111.03606}

\bibitem[{Abbott {et~al.}(2021{\natexlab{b}})}]{LIGOScientific:2021iyk}
---. 2021{\natexlab{b}}.
\newblock \doarXiv{2111.03608}

\bibitem[{Abbott {et~al.}(2021{\natexlab{c}})}]{LIGOScientific:2020lst}
---. 2021{\natexlab{c}}, Astrophys. J., 915, 86,
  \dodoi{10.3847/1538-4357/abee15}

\bibitem[{Acernese {et~al.}(2015)}]{TheVirgo:2014hva}
Acernese, F., {et~al.} 2015, Class. Quantum Grav., 32, 024001,
  \dodoi{10.1088/0264-9381/32/2/024001}

\bibitem[{Allen {et~al.}(2012)Allen, Anderson, Brady, Brown, \&
  Creighton}]{Allen:2005fk}
Allen, B., Anderson, W.~G., Brady, P.~R., Brown, D.~A., \& Creighton, J. D.~E.
  2012, Phys. Rev. D, 85, 122006, \dodoi{10.1103/PhysRevD.85.122006}

\bibitem[{Amiri {et~al.}(2018)}]{CHIMEFRB:2018mlh}
Amiri, M., {et~al.} 2018, \dodoi{10.3847/1538-4357/aad188}

\bibitem[{Amiri {et~al.}(2021)}]{CHIMEFRB:2021srp}
---. 2021, Astrophys. J. Supp., 257, 59, \dodoi{10.3847/1538-4365/ac33ab}

\bibitem[{Andersen {et~al.}(2020)}]{CHIMEFRB:2020abu}
Andersen, B.~C., {et~al.} 2020, Nature, 587, 54,
  \dodoi{10.1038/s41586-020-2863-y}

\bibitem[{{Ashton} {et~al.}(2018){Ashton}, {Burns}, {Dal Canton}, {Dent},
  {Eggenstein}, {Nielsen}, {Prix}, {Was}, \& {Zhu}}]{2018ApJ...860....6A}
{Ashton}, G., {Burns}, E., {Dal Canton}, T., {et~al.} 2018, \apj, 860, 6,
  \dodoi{10.3847/1538-4357/aabfd2}

\bibitem[{Biwer {et~al.}(2019)Biwer, Capano, De, Cabero, Brown, Nitz, \&
  Raymond}]{Biwer:2018osg}
Biwer, C.~M., Capano, C.~D., De, S., {et~al.} 2019, Publ. Astron. Soc. Pac.,
  131, 024503, \dodoi{10.1088/1538-3873/aaef0b}

\bibitem[{Blanchet(2002)}]{Blanchet:2002av}
Blanchet, L. 2002, Living Rev. Rel., 5, 3

\bibitem[{Bochenek {et~al.}(2020)Bochenek, Ravi, Belov, Hallinan, Kocz,
  Kulkarni, \& McKenna}]{Bochenek:2020zxn}
Bochenek, C.~D., Ravi, V., Belov, K.~V., {et~al.} 2020, Nature, 587, 59,
  \dodoi{10.1038/s41586-020-2872-x}

\bibitem[{Brown {et~al.}(2012)Brown, Harry, Lundgren, \& Nitz}]{Brown:2012qf}
Brown, D.~A., Harry, I., Lundgren, A., \& Nitz, A.~H. 2012, Phys. Rev. D, 86,
  084017, \dodoi{10.1103/PhysRevD.86.084017}

\bibitem[{Callister {et~al.}(2016)Callister, Kanner, \&
  Weinstein}]{Callister:2016vtl}
Callister, T., Kanner, J., \& Weinstein, A. 2016, Astrophys. J. Lett., 825,
  L12, \dodoi{10.3847/2041-8205/825/1/L12}

\bibitem[{Capano {et~al.}(2020)Capano, Tews, Brown, Margalit, De, Kumar, Brown,
  Krishnan, \& Reddy}]{Capano:2019eae}
Capano, C.~D., Tews, I., Brown, S.~M., {et~al.} 2020, Nature Astron., 4, 625,
  \dodoi{10.1038/s41550-020-1014-6}

\bibitem[{Davies {et~al.}(2020)Davies, Dent, T\'apai, Harry, McIsaac, \&
  Nitz}]{Davies:2020tsx}
Davies, G.~S., Dent, T., T\'apai, M., {et~al.} 2020, Phys. Rev. D, 102, 022004,
  \dodoi{10.1103/PhysRevD.102.022004}

\bibitem[{Dietrich {et~al.}(2020)Dietrich, Coughlin, Pang, Bulla, Heinzel,
  Issa, Tews, \& Antier}]{Dietrich:2020efo}
Dietrich, T., Coughlin, M.~W., Pang, P. T.~H., {et~al.} 2020, Science, 370,
  1450, \dodoi{10.1126/science.abb4317}

\bibitem[{Droz {et~al.}(1999)Droz, Knapp, Poisson, \& Owen}]{Droz:1999qx}
Droz, S., Knapp, D.~J., Poisson, E., \& Owen, B.~J. 1999, \prd, 59, 124016

\bibitem[{Eichler {et~al.}(1989)Eichler, Livio, Piran, \& Schramm}]{cite-key}
Eichler, D., Livio, M., Piran, T., \& Schramm, D.~N. 1989, Nature, 340, 126,
  \dodoi{10.1038/340126a0}

\bibitem[{Faye {et~al.}(2012)Faye, Marsat, Blanchet, \& Iyer}]{Faye:2012we}
Faye, G., Marsat, S., Blanchet, L., \& Iyer, B.~R. 2012, Class. Quant. Grav.,
  29, 175004, \dodoi{10.1088/0264-9381/29/17/175004}

\bibitem[{Fishbach {et~al.}(2019)}]{LIGOScientific:2018gmd}
Fishbach, M., {et~al.} 2019, Astrophys. J. Lett., 871, L13,
  \dodoi{10.3847/2041-8213/aaf96e}

\bibitem[{Geng \& Huang(2015)}]{Geng:2015vza}
Geng, J.~J., \& Huang, Y.~F. 2015, Astrophys. J., 809, 24,
  \dodoi{10.1088/0004-637X/809/1/24}

\bibitem[{{Goodman}(1986)}]{1986ApJ...308L..47G}
{Goodman}, J. 1986, \apjl, 308, L47, \dodoi{10.1086/184741}

\bibitem[{Guidorzi {et~al.}(2017)}]{Guidorzi:2017ogy}
Guidorzi, C., {et~al.} 2017, Astrophys. J. Lett., 851, L36,
  \dodoi{10.3847/2041-8213/aaa009}

\bibitem[{Harry {et~al.}(2009)Harry, Allen, \& Sathyaprakash}]{Harry:2009ea}
Harry, I.~W., Allen, B., \& Sathyaprakash, B. 2009, Phys. Rev. D, 80, 104014,
  \dodoi{10.1103/PhysRevD.80.104014}

\bibitem[{Hotokezaka {et~al.}(2019)Hotokezaka, Nakar, Gottlieb, Nissanke,
  Masuda, Hallinan, Mooley, \& Deller}]{Hotokezaka:2018dfi}
Hotokezaka, K., Nakar, E., Gottlieb, O., {et~al.} 2019, Nature Astron., 3, 940,
  \dodoi{10.1038/s41550-019-0820-1}

\bibitem[{James {et~al.}(2019)James, Anderson, Wen, Bosveld, Chu, Kovalam,
  Slaven-Blair, \& Williams}]{James:2019xca}
James, C.~W., Anderson, G.~E., Wen, L., {et~al.} 2019, Mon. Not. Roy. Astron.
  Soc., 489, L75, \dodoi{10.1093/mnrasl/slz129}

\bibitem[{{Keane} {et~al.}(2012){Keane}, {Stappers}, {Kramer}, \&
  {Lyne}}]{2012MNRAS.425L..71K}
{Keane}, E.~F., {Stappers}, B.~W., {Kramer}, M., \& {Lyne}, A.~G. 2012, \mnras,
  425, L71, \dodoi{10.1111/j.1745-3933.2012.01306.x}

\bibitem[{Landry {et~al.}(2020)Landry, Essick, \&
  Chatziioannou}]{Landry:2020vaw}
Landry, P., Essick, R., \& Chatziioannou, K. 2020, Phys. Rev. D, 101, 123007,
  \dodoi{10.1103/PhysRevD.101.123007}

\bibitem[{Li {et~al.}(2020)}]{Insight-HXMTTeam:2020dmu}
Li, C.~K., {et~al.} 2020.
\newblock \doarXiv{2005.11071}

\bibitem[{Liu {et~al.}(2016)Liu, Romero, Liu, \& Li}]{Liu:2016olx}
Liu, T., Romero, G.~E., Liu, M.-L., \& Li, A. 2016, Astrophys. J., 826, 82,
  \dodoi{10.3847/0004-637X/826/1/82}

\bibitem[{{Lorimer} {et~al.}(2007){Lorimer}, {Bailes}, {McLaughlin},
  {Narkevic}, \& {Crawford}}]{2007Sci...318..777L}
{Lorimer}, D.~R., {Bailes}, M., {McLaughlin}, M.~A., {Narkevic}, D.~J., \&
  {Crawford}, F. 2007, Science, 318, 777, \dodoi{10.1126/science.1147532}

\bibitem[{Margalit {et~al.}(2020)Margalit, Beniamini, Sridhar, \&
  Metzger}]{Margalit_2020}
Margalit, B., Beniamini, P., Sridhar, N., \& Metzger, B.~D. 2020, The
  Astrophysical Journal, 899, L27, \dodoi{10.3847/2041-8213/abac57}

\bibitem[{{Meegan} {et~al.}(2009){Meegan}, {Lichti}, {Bhat}, {Bissaldi},
  {Briggs}, {Connaughton}, {Diehl}, {Fishman}, {Greiner}, {Hoover}, {van der
  Horst}, {von Kienlin}, {Kippen}, {Kouveliotou}, {McBreen}, {Paciesas},
  {Preece}, {Steinle}, {Wallace}, {Wilson}, \&
  {Wilson-Hodge}}]{2009ApJ...702..791M}
{Meegan}, C., {Lichti}, G., {Bhat}, P.~N., {et~al.} 2009, \apj, 702, 791,
  \dodoi{10.1088/0004-637X/702/1/791}

\bibitem[{Mereghetti {et~al.}(2020)}]{Mereghetti:2020unm}
Mereghetti, S., {et~al.} 2020, Astrophys. J. Lett., 898, L29,
  \dodoi{10.3847/2041-8213/aba2cf}

\bibitem[{Metzger {et~al.}(2017)Metzger, Berger, \& Margalit}]{Metzger_2017}
Metzger, B.~D., Berger, E., \& Margalit, B. 2017, The Astrophysical Journal,
  841, 14, \dodoi{10.3847/1538-4357/aa633d}

\bibitem[{Metzger {et~al.}(2019)Metzger, Margalit, \&
  Sironi}]{10.1093/mnras/stz700}
Metzger, B.~D., Margalit, B., \& Sironi, L. 2019, Monthly Notices of the Royal
  Astronomical Society, 485, 4091, \dodoi{10.1093/mnras/stz700}

\bibitem[{{Narayan} {et~al.}(1992){Narayan}, {Paczynski}, \&
  {Piran}}]{1992ApJ...395L..83N}
{Narayan}, R., {Paczynski}, B., \& {Piran}, T. 1992, \apjl, 395, L83,
  \dodoi{10.1086/186493}

\bibitem[{Nitz {et~al.}(2017)Nitz, Dent, Dal~Canton, Fairhurst, \&
  Brown}]{Nitz:2017svb}
Nitz, A.~H., Dent, T., Dal~Canton, T., Fairhurst, S., \& Brown, D.~A. 2017,
  Astrophys. J., 849, 118, \dodoi{10.3847/1538-4357/aa8f50}

\bibitem[{Nitz {et~al.}(2020)Nitz, Dent, Davies, \& Harry}]{Nitz:2020naa}
Nitz, A.~H., Dent, T., Davies, G.~S., \& Harry, I. 2020, Astrophys. J., 897,
  169, \dodoi{10.3847/1538-4357/ab96c7}

\bibitem[{Nitz {et~al.}(2021)Nitz, Kumar, Wang, Kastha, Wu, Sch\"afer,
  Dhurkunde, \& Capano}]{Nitz:2021zwj}
Nitz, A.~H., Kumar, S., Wang, Y.-F., {et~al.} 2021.
\newblock \doarXiv{2112.06878}

\bibitem[{Nitz {et~al.}(2019)Nitz, Nielsen, \& Capano}]{Nitz:2019bxt}
Nitz, A.~H., Nielsen, A.~B., \& Capano, C.~D. 2019, Astrophys. J. Lett., 876,
  L4, \dodoi{10.3847/2041-8213/ab18a1}

\bibitem[{Nitz {et~al.}(2018)Nitz, Harry, Willis, Biwer, Brown, Pekowsky,
  Dal~Canton, Williamson, Dent, Capano, Massinger, Lenon, Nielsen, \&
  Cabero}]{pycbc-github}
Nitz, A.~H., Harry, I.~W., Willis, J.~L., {et~al.} 2018, {PyCBC Software},
  \url{https://github.com/gwastro/pycbc},  GitHub

\bibitem[{Niu {et~al.}(2021)}]{Niu:2021xug}
Niu, C.-H., {et~al.} 2021, Astrophys. J. Lett., 909, L8,
  \dodoi{10.3847/2041-8213/abe7f0}

\bibitem[{Olsen {et~al.}(2022)Olsen, Venumadhav, Mushkin, Roulet, Zackay, \&
  Zaldarriaga}]{Olsen:2022pin}
Olsen, S., Venumadhav, T., Mushkin, J., {et~al.} 2022.
\newblock \doarXiv{2201.02252}

\bibitem[{{Paczynski}(1986)}]{1986ApJ...308L..43P}
{Paczynski}, B. 1986, \apjl, 308, L43, \dodoi{10.1086/184740}

\bibitem[{Popov \& Postnov(2007)}]{Popov:2007uv}
Popov, S.~B., \& Postnov, K.~A. 2007.
\newblock \doarXiv{0710.2006}

\bibitem[{Ridnaia {et~al.}(2021)}]{Ridnaia:2020gcv}
Ridnaia, A., {et~al.} 2021, Nature Astron., 5, 372,
  \dodoi{10.1038/s41550-020-01265-0}

\bibitem[{Sathyaprakash \& Dhurandhar(1991)}]{Sathyaprakash:1991mt}
Sathyaprakash, B.~S., \& Dhurandhar, S.~V. 1991, Phys. Rev. D, 44, 3819,
  \dodoi{10.1103/PhysRevD.44.3819}

\bibitem[{Singer \& Price(2016)}]{PhysRevD.93.024013}
Singer, L.~P., \& Price, L.~R. 2016, Phys. Rev. D, 93, 024013,
  \dodoi{10.1103/PhysRevD.93.024013}

\bibitem[{Speagle(2020)}]{speagle:2019}
Speagle, J.~S. 2020, Monthly Notices of the Royal Astronomical Society, 493,
  3132, \dodoi{10.1093/mnras/staa278}

\bibitem[{{Spitler} {et~al.}(2016){Spitler}, {Scholz}, {Hessels}, {Bogdanov},
  {Brazier}, {Camilo}, {Chatterjee}, {Cordes}, {Crawford}, {Deneva}, {Ferdman},
  {Freire}, {Kaspi}, {Lazarus}, {Lynch}, {Madsen}, {McLaughlin}, {Patel},
  {Ransom}, {Seymour}, {Stairs}, {Stappers}, {van Leeuwen}, \&
  {Zhu}}]{2016Natur.531..202S}
{Spitler}, L.~G., {Scholz}, P., {Hessels}, J.~W.~T., {et~al.} 2016, \nat, 531,
  202, \dodoi{10.1038/nature17168}

\bibitem[{Sridhar {et~al.}(2021)Sridhar, Zrake, Metzger, Sironi, \&
  Giannios}]{Sridhar:2020uez}
Sridhar, N., Zrake, J., Metzger, B.~D., Sironi, L., \& Giannios, D. 2021, Mon.
  Not. Roy. Astron. Soc., 501, 3184, \dodoi{10.1093/mnras/staa3794}

\bibitem[{Tavani {et~al.}(2021)}]{Tavani:2020adq}
Tavani, M., {et~al.} 2021, Nature Astron., 5, 401,
  \dodoi{10.1038/s41550-020-01276-x}

\bibitem[{{The CHIME/FRB Collaboration}(2018)}]{FRB20190701E}
{The CHIME/FRB Collaboration}. 2018, {FRB 20190701E},
  \url{https://www.chime-frb.ca/catalog/FRB20190701E}

\bibitem[{{The LIGO Scientific Collaboration} {et~al.}(2022){The LIGO
  Scientific Collaboration}, {the Virgo Collaboration}, {the KAGRA
  Collaboration}, {the CHIME/FRB Collaboration}, {:}, {Abbott}, {Abbott},
  {Acernese}, {Ackley}, {Adams}, {Adhikari}, {Adhikari}, {Adya}, {Affeldt},
  {Agarwal}, {Agathos}, {Agatsuma}, {Aggarwal}, {Aguiar}, {Aiello}, {Ain},
  {Ajith}, {Akutsu}, {Albanesi}, {Allocca}, {Altin}, {Amato}, {Anand}, {Anand},
  {Ananyeva}, {Anderson}, {Anderson}, {Ando}, {Andrade}, {Andres},
  {Andri{\'c}}, {Angelova}, {Ansoldi}, {Antelis}, {Antier}, {Appert}, {Arai},
  {Arai}, {Arai}, {Araki}, {Araya}, {Araya}, {Areeda}, {Ar{\`e}ne}, {Aritomi},
  {Arnaud}, {Aronson}, {Arun}, {Asada}, {Asali}, {Ashton}, {Aso}, {Assiduo},
  {Aston}, {Astone}, {Aubin}, {Austin}, {Babak}, {Badaracco}, {Bader},
  {Badger}, {Bae}, {Bae}, {Baer}, {Bagnasco}, {Bai}, {Baiotti}, {Baird},
  {Bajpai}, {Ball}, {Ballardin}, {Ballmer}, {Balsamo}, {Baltus}, {Banagiri},
  {Bankar}, {Barayoga}, {Barbieri}, {Barish}, {Barker}, {Barneo}, {Barone},
  {Barr}, {Barsotti}, {Barsuglia}, {Barta}, {Bartlett}, {Barton}, {Bartos},
  {Bassiri}, {Basti}, {Bawaj}, {Bayley}, {Baylor}, {Bazzan}, {B{\'e}csy},
  {Bedakihale}, {Bejger}, {Belahcene}, {Benedetto}, {Beniwal}, {Bennett},
  {Bentley}, {BenYaala}, {Bergamin}, {Berger}, {Bernuzzi}, {Berry},
  {Bersanetti}, {Bertolini}, {Betzwieser}, {Beveridge}, {Bhandare}, {Bhardwaj},
  {Bhattacharjee}, {Bhaumik}, {Bilenko}, {Billingsley}, {Bini}, {Birney},
  {Birnholtz}, {Biscans}, {Bischi}, {Biscoveanu}, {Bisht}, {Biswas}, {Bitossi},
  {Bizouard}, {Blackburn}, {Blair}, {Blair}, {Blair}, {Bobba}, {Bode}, {Boer},
  {Bogaert}, {Boldrini}, {Bonavena}, {Bondu}, {Bonilla}, {Bonnand}, {Booker},
  {Boom}, {Bork}, {Boschi}, {Bose}, {Bose}, {Bossilkov}, {Boudart},
  {Bouffanais}, {Bozzi}, {Bradaschia}, {Brady}, {Bramley}, {Branch},
  {Branchesi}, {Brau}, {Breschi}, {Briant}, {Briggs}, {Brillet}, {Brinkmann},
  {Brockill}, {Brooks}, {Brooks}, {Brown}, {Brunett}, {Bruno}, {Bruntz},
  {Bryant}, {Buchanan}, {Bulik}, {Bulten}, {Buonanno}, {Buscicchio},
  {Buskulic}, {Buy}, {Byer}, {Cadonati}, {Cagnoli}, {Cahillane}, {Calder{\'o}n
  Bustillo}, {Callaghan}, {Callister}, {Calloni}, {Cameron}, {Camp}, {Canepa},
  {Canevarolo}, {Cannavacciuolo}, {Cannon}, {Cao}, {Cao}, {Capocasa}, {Capote},
  {Carapella}, {Carbognani}, {Carlin}, {Carney}, {Carpinelli}, {Carrillo},
  {Carullo}, {Carver}, {Casanueva Diaz}, {Casentini}, {Castaldi}, {Caudill},
  {Cavagli{\`a}}, {Cavalier}, {Cavalieri}, {Ceasar}, {Cella},
  {Cerd{\'a}-Dur{\'a}n}, {Cesarini}, {Chaibi}, {Chakravarti}, {Chalathadka
  Subrahmanya}, {Champion}, {Chan}, {Chan}, {Chan}, {Chan}, {Chan}, {Chandra},
  {Chanial}, {Chao}, {Charlton}, {Chase}, {Chassande-Mottin}, {Chatterjee},
  {Chatterjee}, {Chatterjee}, {Chaturvedi}, {Chaty}, {Chen}, {Chen}, {Chen},
  {Chen}, {Chen}, {Chen}, {Chen}, {Chen}, {Cheng}, {Cheong}, {Cheung}, {Chia},
  {Chiadini}, {Chiang}, {Chiarini}, {Chierici}, {Chincarini}, {Chiofalo},
  {Chiummo}, {Cho}, {Cho}, {Choudhary}, {Choudhary}, {Christensen}, {Chu},
  {Chu}, {Chu}, {Chua}, {Chung}, {Ciani}, {Ciecielag}, {Cie{\'s}lar},
  {Cifaldi}, {Ciobanu}, {Ciolfi}, {Cipriano}, {Cirone}, {Clara}, {Clark},
  {Clark}, {Clarke}, {Clearwater}, {Clesse}, {Cleva}, {Coccia}, {Codazzo},
  {Cohadon}, {Cohen}, {Cohen}, {Colleoni}, {Collette}, {Colombo}, {Colpi},
  {Compton}, {Constancio}, {Conti}, {Cooper}, {Corban}, {Corbitt},
  {Cordero-Carri{\'o}n}, {Corezzi}, {Corley}, {Cornish}, {Corre}, {Corsi},
  {Cortese}, {Costa}, {Cotesta}, {Coughlin}, {Coulon}, {Countryman}, {Cousins},
  {Couvares}, {Coward}, {Cowart}, {Coyne}, {Coyne}, {Creighton}, {Creighton},
  {Criswell}, {Croquette}, {Crowder}, {Cudell}, {Cullen}, {Cumming},
  {Cummings}, {Cunningham}, {Cuoco}, {Cury{\l}o}, {Dabadie}, {Dal Canton},
  {Dall'Osso}, {D{\'a}lya}, {Dana}, {DaneshgaranBajastani}, {D'Angelo},
  {Danilishin}, {D'Antonio}, {Danzmann}, {Darsow-Fromm}, {Dasgupta}, {Datrier},
  {Datta}, {Dattilo}, {Dave}, {Davier}, {Davies}, {Davis}, {Davis}, {Daw},
  {Dean}, {DeBra}, {Deenadayalan}, {Degallaix}, {De Laurentis},
  {Del{\'e}glise}, {Del Favero}, {De Lillo}, {De Lillo}, {Del Pozzo},
  {DeMarchi}, {De Matteis}, {D'Emilio}, {Demos}, {Dent}, {Depasse}, {De
  Pietri}, {De Rosa}, {De Rossi}, {DeSalvo}, {De Simone}, {Dhurandhar},
  {D{\'\i}az}, {Diaz-Ortiz}, {Didio}, {Dietrich}, {Di Fiore}, {Di Fronzo}, {Di
  Giorgio}, {Di Giovanni}, {Di Giovanni}, {Di Girolamo}, {Di Lieto}, {Ding},
  {Di Pace}, {Di Palma}, {Di Renzo}, {Divakarla}, {Dmitriev}, {Doctor},
  {D'Onofrio}, {Donovan}, {Dooley}, {Doravari}, {Dorrington}, {Drago},
  {Driggers}, {Drori}, {Ducoin}, {Dupej}, {Durante}, {D'Urso}, {Duverne},
  {Dwyer}, {Eassa}, {Easter}, {Ebersold}, {Eckhardt}, {Eddolls}, {Edelman},
  {Edo}, {Edy}, {Effler}, {Eguchi}, {Eichholz}, {Eikenberry}, {Eisenmann},
  {Eisenstein}, {Ejlli}, {Engelby}, {Enomoto}, {Errico}, {Essick},
  {Estell{\'e}s}, {Estevez}, {Etienne}, {Etzel}, {Evans}, {Evans}, {Ewing},
  {Fafone}, {Fair}, {Fairhurst}, {Farah}, {Farinon}, {Farr}, {Farr}, {Farrow},
  {Fauchon-Jones}, {Favaro}, {Favata}, {Fays}, {Fazio}, {Feicht}, {Fejer},
  {Fenyvesi}, {Ferguson}, {Fernandez-Galiana}, {Ferrante}, {Ferreira},
  {Fidecaro}, {Figura}, {Fiori}, {Fishbach}, {Fisher}, {Fittipaldi}, {Fiumara},
  {Flaminio}, {Floden}, {Fong}, {Font}, {Fornal}, {Forsyth}, {Franke},
  {Frasca}, {Frasconi}, {Frederick}, {Freed}, {Frei}, {Freise}, {Frey},
  {Fritschel}, {Frolov}, {Fronz{\'e}}, {Fujii}, {Fujikawa}, {Fukunaga},
  {Fukushima}, {Fulda}, {Fyffe}, {Gabbard}, {Gadre}, {Gair}, {Gais},
  {Galaudage}, {Gamba}, {Ganapathy}, {Ganguly}, {Gao}, {Gaonkar}, {Garaventa},
  {Garc{\'\i}a-N{\'u}{\~n}ez}, {Garc{\'\i}a-Quir{\'o}s}, {Garufi}, {Gateley},
  {Gaudio}, {Gayathri}, {Ge}, {Gemme}, {Gennai}, {George}, {Gerberding},
  {Gergely}, {Gewecke}, {Ghonge}, {Ghosh}, {Ghosh}, {Ghosh}, {Ghosh},
  {Giacomazzo}, {Giacoppo}, {Giaime}, {Giardina}, {Gibson}, {Gier}, {Giesler},
  {Giri}, {Gissi}, {Glanzer}, {Gleckl}, {Godwin}, {Goetz}, {Goetz}, {Gohlke},
  {Goncharov}, {Gonz{\'a}lez}, {Gopakumar}, {Gosselin}, {Gouaty}, {Gould},
  {Grace}, {Grado}, {Granata}, {Granata}, {Grant}, {Gras}, {Grassia}, {Gray},
  {Gray}, {Greco}, {Green}, {Green}, {Gretarsson}, {Gretarsson}, {Griffith},
  {Griffiths}, {Griggs}, {Grignani}, {Grimaldi}, {Grimm}, {Grote}, {Grunewald},
  {Gruning}, {Guerra}, {Guidi}, {Guimaraes}, {Guix{\'e}}, {Gulati}, {Guo},
  {Guo}, {Gupta}, {Gupta}, {Gupta}, {Gustafson}, {Gustafson}, {Guzman}, {Ha},
  {Haegel}, {Hagiwara}, {Haino}, {Halim}, {Hall}, {Hamilton}, {Hammond}, {Han},
  {Haney}, {Hanks}, {Hanna}, {Hannam}, {Hannuksela}, {Hansen}, {Hansen},
  {Hanson}, {Harder}, {Hardwick}, {Haris}, {Harms}, {Harry}, {Harry},
  {Hartwig}, {Hasegawa}, {Haskell}, {Hasskew}, {Haster}, {Hattori}, {Haughian},
  {Hayakawa}, {Hayama}, {Hayes}, {Healy}, {Heidmann}, {Heidt}, {Heintze},
  {Heinze}, {Heinzel}, {Heitmann}, {Hellman}, {Hello}, {Helmling-Cornell},
  {Hemming}, {Hendry}, {Heng}, {Hennes}, {Hennig}, {Hennig}, {Hernandez},
  {Hernandez Vivanco}, {Heurs}, {Hild}, {Hill}, {Himemoto}, {Hines},
  {Hiranuma}, {Hirata}, {Hirose}, {Hochheim}, {Hofman}, {Hohmann}, {Holcomb},
  {Holland}, {Hollows}, {Holmes}, {Holt}, {Holz}, {Hong}, {Hopkins}, {Hough},
  {Hourihane}, {Howell}, {Hoy}, {Hoyland}, {Hreibi}, {Hsieh}, {Hsu}, {Huang},
  {Huang}, {Huang}, {Huang}, {Huang}, {Huang}, {H{\"u}bner}, {Huddart},
  {Hughey}, {Hui}, {Hui}, {Husa}, {Huttner}, {Huxford}, {Huynh-Dinh}, {Ide},
  {Idzkowski}, {Iess}, {Ikenoue}, {Imam}, {Inayoshi}, {Ingram}, {Inoue},
  {Ioka}, {Isi}, {Isleif}, {Ito}, {Itoh}, {Iyer}, {Izumi}, {JaberianHamedan},
  {Jacqmin}, {Jadhav}, {Jadhav}, {James}, {Jan}, {Jani}, {Janquart},
  {Janssens}, {Janthalur}, {Jaranowski}, {Jariwala}, {Jaume}, {Jenkins},
  {Jenner}, {Jeon}, {Jeunon}, {Jia}, {Jin}, {Johns}, {Jones}, {Jones}, {Jones},
  {Jones}, {Jones}, {Jonker}, {Ju}, {Jung}, {Jung}, {Junker}, {Juste},
  {Kaihotsu}, {Kajita}, {Kakizaki}, {Kalaghatgi}, {Kalogera}, {Kamai},
  {Kamiizumi}, {Kanda}, {Kandhasamy}, {Kang}, {Kanner}, {Kao}, {Kapadia},
  {Kapasi}, {Karat}, {Karathanasis}, {Karki}, {Kashyap}, {Kasprzack},
  {Kastaun}, {Katsanevas}, {Katsavounidis}, {Katzman}, {Kaur}, {Kawabe},
  {Kawaguchi}, {Kawai}, {Kawasaki}, {K{\'e}f{\'e}lian}, {Keitel}, {Key},
  {Khadka}, {Khalili}, {Khan}, {Khazanov}, {Khetan}, {Khursheed}, {Kijbunchoo},
  {Kim}, {Kim}, {Kim}, {Kim}, {Kim}, {Kim}, {Kimball}, {Kimura},
  {Kinley-Hanlon}, {Kirchhoff}, {Kissel}, {Kita}, {Kitazawa}, {Kleybolte},
  {Klimenko}, {Knee}, {Knowles}, {Knyazev}, {Koch}, {Koekoek}, {Kojima},
  {Kokeyama}, {Koley}, {Kolitsidou}, {Kolstein}, {Komori}, {Kondrashov},
  {Kong}, {Kontos}, {Koper}, {Korobko}, {Kotake}, {Kovalam}, {Kozak},
  {Kozakai}, {Kozu}, {Kringel}, {Krishnendu}, {Kr{\'o}lak}, {Kuehn}, {Kuei},
  {Kuijer}, {Kumar}, {Kumar}, {Kumar}, {Kumar}, {Kume}, {Kuns}, {Kuo}, {Kuo},
  {Kuromiya}, {Kuroyanagi}, {Kusayanagi}, {Kuwahara}, {Kwak}, {Lagabbe},
  {Laghi}, {Lalande}, {Lam}, {Lamberts}, {Landry}, {Lane}, {Lang}, {Lange},
  {Lantz}, {La Rosa}, {Lartaux-Vollard}, {Lasky}, {Laxen}, {Lazzarini},
  {Lazzaro}, {Leaci}, {Leavey}, {Lecoeuche}, {Lee}, {Lee}, {Lee}, {Lee}, {Lee},
  {Lee}, {Lehmann}, {Lema{\^\i}tre}, {Leonardi}, {Leroy}, {Letendre},
  {Levesque}, {Levin}, {Leviton}, {Leyde}, {Li}, {Li}, {Li}, {Li}, {Li}, {Li},
  {Lin}, {Lin}, {Lin}, {Lin}, {Lin}, {Linde}, {Linker}, {Linley}, {Littenberg},
  {Liu}, {Liu}, {Liu}, {Liu}, {Llamas}, {Llorens-Monteagudo}, {Lo}, {Lockwood},
  {London}, {Longo}, {Lopez}, {Lopez Portilla}, {Lorenzini}, {Loriette},
  {Lormand}, {Losurdo}, {Lott}, {Lough}, {Lousto}, {Lovelace}, {Lucaccioni},
  {L{\"u}ck}, {Lumaca}, {Lundgren}, {Luo}, {Lynam}, {Macas}, {MacInnis},
  {Macleod}, {MacMillan}, {Macquet}, {Maga{\~n}a Hernandez}, {Magazz{\`u}},
  {Magee}, {Maggiore}, {Magnozzi}, {Mahesh}, {Majorana}, {Makarem},
  {Maksimovic}, {Maliakal}, {Malik}, {Man}, {Mandic}, {Mangano}, {Mango},
  {Mansell}, {Manske}, {Mantovani}, {Mapelli}, {Marchesoni}, {Marchio},
  {Marion}, {Mark}, {M{\'a}rka}, {M{\'a}rka}, {Markakis}, {Markosyan},
  {Markowitz}, {Maros}, {Marquina}, {Marsat}, {Martelli}, {Martin}, {Martin},
  {Martinez}, {Martinez}, {Martinez}, {Martinovic}, {Martynov}, {Marx},
  {Masalehdan}, {Mason}, {Massera}, {Masserot}, {Massinger}, {Masso-Reid},
  {Mastrogiovanni}, {Matas}, {Mateu-Lucena}, {Matichard}, {Matiushechkina},
  {Mavalvala}, {McCann}, {McCarthy}, {McClelland}, {McClincy}, {McCormick},
  {McCuller}, {McGhee}, {McGuire}, {McIsaac}, {McIver}, {McRae}, {McWilliams},
  {Meacher}, {Mehmet}, {Mehta}, {Meijer}, {Melatos}, {Melchor}, {Mendell},
  {Menendez-Vazquez}, {Menoni}, {Mercer}, {Mereni}, {Merfeld}, {Merilh},
  {Merritt}, {Merzougui}, {Meshkov}, {Messenger}, {Messick}, {Meyers},
  {Meylahn}, {Mhaske}, {Miani}, {Miao}, {Michaloliakos}, {Michel}, {Michimura},
  {Middleton}, {Milano}, {Miller}, {Miller}, {Miller}, {Millhouse}, {Mills},
  {Milotti}, {Minazzoli}, {Minenkov}, {Mio}, {Mir}, {Miravet-Ten{\'e}s},
  {Mishra}, {Mishra}, {Mistry}, {Mitra}, {Mitrofanov}, {Mitselmakher},
  {Mittleman}, {Miyakawa}, {Miyamoto}, {Miyazaki}, {Miyo}, {Miyoki}, {Mo},
  {Moguel}, {Mogushi}, {Mohapatra}, {Mohite}, {Molina}, {Molina-Ruiz},
  {Mondin}, {Montani}, {Moore}, {Moraru}, {Morawski}, {More}, {Moreno},
  {Moreno}, {Mori}, {Morisaki}, {Moriwaki}, {Mours}, {Mow-Lowry}, {Mozzon},
  {Muciaccia}, {Mukherjee}, {Mukherjee}, {Mukherjee}, {Mukherjee}, {Mukherjee},
  {Mukund}, {Mullavey}, {Munch}, {Mu{\~n}iz}, {Murray}, {Musenich}, {Muusse},
  {Nadji}, {Nagano}, {Nagano}, {Nagar}, {Nakamura}, {Nakano}, {Nakano},
  {Nakashima}, {Nakayama}, {Napolano}, {Nardecchia}, {Narikawa}, {Naticchioni},
  {Nayak}, {Nayak}, {Negishi}, {Neil}, {Neilson}, {Nelemans}, {Nelson}, {Nery},
  {Neubauer}, {Neunzert}, {Ng}, {Ng}, {Nguyen}, {Nguyen}, {Nguyen}, {Nguyen
  Quynh}, {Ni}, {Nichols}, {Nishizawa}, {Nissanke}, {Nitoglia}, {Nocera},
  {Norman}, {North}, {Nozaki}, {Nuttall}, {Oberling}, {O'Brien}, {Obuchi},
  {O'Dell}, {Oelker}, {Ogaki}, {Oganesyan}, {Oh}, {Oh}, {Oh}, {Ohashi},
  {Ohishi}, {Ohkawa}, {Ohme}, {Ohta}, {Okada}, {Okutani}, {Okutomi},
  {Olivetto}, {Oohara}, {Ooi}, {Oram}, {O'Reilly}, {Ormiston}, {Ormsby},
  {Ortega}, {O'Shaughnessy}, {O'Shea}, {Oshino}, {Ossokine}, {Osthelder},
  {Otabe}, {Ottaway}, {Overmier}, {Pace}, {Pagano}, {Page}, {Pagliaroli},
  {Pai}, {Pai}, {Palamos}, {Palashov}, {Palomba}, {Pan}, {Pan}, {Panda},
  {Pang}, {Pang}, {Pankow}, {Pannarale}, {Pant}, {Panther}, {Paoletti},
  {Paoli}, {Paolone}, {Parisi}, {Park}, {Park}, {Parker}, {Pascucci},
  {Pasqualetti}, {Passaquieti}, {Passuello}, {Patel}, {Pathak}, {Patricelli},
  {Patron}, {Patrone}, {Paul}, {Payne}, {Pedraza}, {Pegoraro}, {Pele},
  {Pe{\~n}a Arellano}, {Penn}, {Perego}, {Pereira}, {Pereira}, {Perez},
  {P{\'e}rigois}, {Perkins}, {Perreca}, {Perri{\`e}s}, {Petermann},
  {Petterson}, {Pfeiffer}, {Pham}, {Phukon}, {Piccinni}, {Pichot},
  {Piendibene}, {Piergiovanni}, {Pierini}, {Pierro}, {Pillant}, {Pillas},
  {Pilo}, {Pinard}, {Pinto}, {Pinto}, {Piotrzkowski}, {Pirello}, {Pitkin},
  {Placidi}, {Planas}, {Plastino}, {Pluchar}, {Poggiani}, {Polini}, {Pong},
  {Ponrathnam}, {Popolizio}, {Porter}, {Poulton}, {Powell}, {Pracchia},
  {Pradier}, {Prajapati}, {Prasai}, {Prasanna}, {Pratten}, {Principe}, {Prodi},
  {Prokhorov}, {Prosposito}, {Prudenzi}, {Puecher}, {Punturo}, {Puosi},
  {Puppo}, {P{\"u}rrer}, {Qi}, {Quetschke}, {Quitzow-James}, {Raab},
  {Raaijmakers}, {Radkins}, {Radulesco}, {Raffai}, {Rail}, {Raja}, {Rajan},
  {Ramirez}, {Ramirez}, {Ramos-Buades}, {Rana}, {Rapagnani}, {Rapol}, {Ray},
  {Raymond}, {Raza}, {Razzano}, {Read}, {Rees}, {Regimbau}, {Rei}, {Reid},
  {Reid}, {Reitze}, {Relton}, {Renzini}, {Rettegno}, {Rezac}, {Ricci},
  {Richards}, {Richardson}, {Richardson}, {Riemenschneider}, {Riles},
  {Rinaldi}, {Rink}, {Rizzo}, {Robertson}, {Robie}, {Robinet}, {Rocchi},
  {Rodriguez}, {Rolland}, {Rollins}, {Romanelli}, {Romano}, {Romel},
  {Romero-Rodr{\'\i}guez}, {Romero-Shaw}, {Romie}, {Ronchini}, {Rosa}, {Rose},
  {Rosi{\'n}ska}, {Ross}, {Rowan}, {Rowlinson}, {Roy}, {Roy}, {Roy}, {Rozza},
  {Ruggi}, {Ryan}, {Sachdev}, {Sadecki}, {Sadiq}, {Sago}, {Saito}, {Saito},
  {Sakai}, {Sakai}, {Sakellariadou}, {Sakuno}, {Salafia}, {Salconi}, {Saleem},
  {Salemi}, {Samajdar}, {Sanchez}, {Sanchez}, {Sanchez}, {Sanchis-Gual},
  {Sanders}, {Sanuy}, {Saravanan}, {Sarin}, {Sassolas}, {Satari}, {Sato},
  {Sato}, {Sauter}, {Savage}, {Sawada}, {Sawant}, {Sawant}, {Sayah},
  {Schaetzl}, {Scheel}, {Scheuer}, {Schiworski}, {Schmidt}, {Schmidt},
  {Schnabel}, {Schneewind}, {Schofield}, {Sch{\"o}nbeck}, {Schulte}, {Schutz},
  {Schwartz}, {Scott}, {Scott}, {Seglar-Arroyo}, {Sekiguchi}, {Sekiguchi},
  {Sellers}, {Sengupta}, {Sentenac}, {Seo}, {Sequino}, {Sergeev}, {Setyawati},
  {Shaffer}, {Shahriar}, {Shams}, {Shao}, {Sharma}, {Sharma}, {Shawhan},
  {Shcheblanov}, {Shibagaki}, {Shikauchi}, {Shimizu}, {Shimoda}, {Shimode},
  {Shinkai}, {Shishido}, {Shoda}, {Shoemaker}, {Shoemaker}, {ShyamSundar},
  {Sieniawska}, {Sigg}, {Singer}, {Singh}, {Singh}, {Singha}, {Sintes},
  {Sipala}, {Skliris}, {Slagmolen}, {Slaven-Blair}, {Smetana}, {Smith},
  {Smith}, {Soldateschi}, {Somala}, {Somiya}, {Son}, {Soni}, {Soni}, {Sordini},
  {Sorrentino}, {Sorrentino}, {Sotani}, {Soulard}, {Souradeep}, {Sowell},
  {Spagnuolo}, {Spencer}, {Spera}, {Srinivasan}, {Srivastava}, {Srivastava},
  {Staats}, {Stachie}, {Steer}, {Steinlechner}, {Steinlechner}, {Stops},
  {Stover}, {Strain}, {Strang}, {Stratta}, {Strunk}, {Sturani}, {Stuver},
  {Sudhagar}, {Sudhir}, {Sugimoto}, {Suh}, {Summerscales}, {Sun}, {Sun},
  {Sunil}, {Sur}, {Suresh}, {Sutton}, {Suzuki}, {Suzuki}, {Swinkels},
  {Szczepa{\'n}czyk}, {Szewczyk}, {Tacca}, {Tagoshi}, {Tait}, {Takahashi},
  {Takahashi}, {Takamori}, {Takano}, {Takeda}, {Takeda}, {Talbot}, {Talbot},
  {Tanaka}, {Tanaka}, {Tanaka}, {Tanaka}, {Tanaka}, {Tanasijczuk}, {Tanioka},
  {Tanner}, {Tao}, {Tao}, {Tapia San Mart{\'\i}n}, {Taranto}, {Tasson},
  {Telada}, {Tenorio}, {Terhune}, {Terkowski}, {Thirugnanasambandam}, {Thomas},
  {Thomas}, {Thompson}, {Thondapu}, {Thorne}, {Thrane}, {Tiwari}, {Tiwari},
  {Tiwari}, {Toivonen}, {Toland}, {Tolley}, {Tomaru}, {Tomigami}, {Tomura},
  {Tonelli}, {Torres-Forn{\'e}}, {Torrie}, {Tosta e Melo}, {T{\"o}yr{\"a}},
  {Trapananti}, {Travasso}, {Traylor}, {Trevor}, {Tringali}, {Tripathee},
  {Troiano}, {Trovato}, {Trozzo}, {Trudeau}, {Tsai}, {Tsai}, {Tsang}, {Tsang},
  {Tsao}, {Tse}, {Tso}, {Tsubono}, {Tsuchida}, {Tsukada}, {Tsuna}, {Tsutsui},
  {Tsuzuki}, {Turbang}, {Turconi}, {Tuyenbayev}, {Ubhi}, {Uchikata},
  {Uchiyama}, {Udall}, {Ueda}, {Uehara}, {Ueno}, {Ueshima}, {Unnikrishnan},
  {Uraguchi}, {Urban}, {Ushiba}, {Utina}, {Vahlbruch}, {Vajente}, {Vajpeyi},
  {Valdes}, {Valentini}, {Valsan}, {van Bakel}, {van Beuzekom}, {van den
  Brand}, {Van Den Broeck}, {Vander-Hyde}, {van der Schaaf}, {van Heijningen},
  {Vanosky}, {van Putten}, {van Remortel}, {Vardaro}, {Vargas}, {Varma},
  {Vas{\'u}th}, {Vecchio}, {Vedovato}, {Veitch}, {Veitch}, {Venneberg},
  {Venugopalan}, {Verkindt}, {Verma}, {Verma}, {Veske}, {Vetrano},
  {Vicer{\'e}}, {Vidyant}, {Viets}, {Vijaykumar}, {Villa-Ortega}, {Vinet},
  {Virtuoso}, {Vitale}, {Vo}, {Vocca}, {von Reis}, {von Wrangel}, {Vorvick},
  {Vyatchanin}, {Wade}, {Wade}, {Wagner}, {Walet}, {Walker}, {Wallace},
  {Wallace}, {Walsh}, {Wang}, {Wang}, {Wang}, {Ward}, {Warner}, {Was},
  {Washimi}, {Washington}, {Watada}, {Watchi}, {Weaver}, {Webster}, {Weinert},
  {Weinstein}, {Weiss}, {Weller}, {Wellmann}, {Wen}, {We{\ss}els}, {Wette},
  {Whelan}, {White}, {Whiting}, {Whittle}, {Wilken}, {Williams}, {Williams},
  {Williamson}, {Willis}, {Willke}, {Wilson}, {Winkler}, {Wipf}, {Wlodarczyk},
  {Woan}, {Woehler}, {Wofford}, {Wong}, {Wu}, {Wu}, {Wu}, {Wu}, {Wysocki},
  {Xiao}, {Xu}, {Yamada}, {Yamamoto}, {Yamamoto}, {Yamamoto}, {Yamamoto},
  {Yamashita}, {Yamazaki}, {Yang}, {Yang}, {Yang}, {Yang}, {Yang}, {Yap},
  {Yeeles}, {Yelikar}, {Ying}, {Yokogawa}, {Yokoyama}, {Yokozawa}, {Yoo},
  {Yoshioka}, {Yu}, {Yu}, {Yuzurihara}, {Zadro{\.z}ny}, {Zanolin}, {Zeidler},
  {Zelenova}, {Zendri}, {Zevin}, {Zhan}, {Zhang}, {Zhang}, {Zhang}, {Zhang},
  {Zhang}, {Zhao}, {Zhao}, {Zhao}, {Zhao}, {Zhou}, {Zhou}, {Zhu}, {Zhu},
  {Zimmerman}, {Zucker}, {Zweizig}, {Bhardwaj}, {Boyle}, {Cassanelli}, {Dong},
  {Fonseca}, {Kaspi}, {Leung}, {Masui}, {Meyers}, {Michilli}, {Ng}, {Pearlman},
  {Petroff}, {Pleunis}, {Rafiei-Ravandi}, {Rahman}, {Ransom}, {Scholz}, {Shin},
  {Smith}, {Stairs}, {Tendulkar}, \& {Zwaniga}}]{2022arXiv220312038T}
{The LIGO Scientific Collaboration}, {the Virgo Collaboration}, {the KAGRA
  Collaboration}, {et~al.} 2022, arXiv e-prints, arXiv:2203.12038.
\newblock \doarXiv{2203.12038}

\bibitem[{{Totani}(2013)}]{2013PASJ...65L..12T}
{Totani}, T. 2013, \pasj, 65, L12, \dodoi{10.1093/pasj/65.5.L12}

\bibitem[{Vachaspati(2008)}]{PhysRevLett.101.141301}
Vachaspati, T. 2008, Phys. Rev. Lett., 101, 141301,
  \dodoi{10.1103/PhysRevLett.101.141301}

\bibitem[{Vallisneri {et~al.}(2015)Vallisneri, Kanner, Williams, Weinstein, \&
  Stephens}]{Vallisneri:2014vxa}
Vallisneri, M., Kanner, J., Williams, R., Weinstein, A., \& Stephens, B. 2015,
  J. Phys. Conf. Ser., 610, 012021, \dodoi{10.1088/1742-6596/610/1/012021}

\bibitem[{Wang {et~al.}(2020)Wang, Wang, Yang, Yu, Zuo, \& Dai}]{Wang:2020aut}
Wang, F.~Y., Wang, Y.~Y., Yang, Y.-P., {et~al.} 2020,
  \dodoi{10.3847/1538-4357/ab74d0}

\bibitem[{Wang {et~al.}(2016)Wang, Yang, Wu, Dai, \& Wang}]{Wang:2016dgs}
Wang, J.-S., Yang, Y.-P., Wu, X.-F., Dai, Z.-G., \& Wang, F.-Y. 2016,
  Astrophys. J. Lett., 822, L7, \dodoi{10.3847/2041-8205/822/1/L7}

\bibitem[{{Wang, Y.-F. and Nitz, A.H.}(2022)}]{datarelease}
{Wang, Y.-F. and Nitz, A.H.} 2022, {Data release associated with this work},
  \url{https://github.com/gwastro/gwfrb-4ogc-chime}

\bibitem[{Zhang(2016)}]{Zhang:2016rli}
Zhang, B. 2016, Astrophys. J. Lett., 827, L31,
  \dodoi{10.3847/2041-8205/827/2/L31}

\bibitem[{Zhang(2020)}]{Zhang:2020qgp}
---. 2020, Nature, 587, 45, \dodoi{10.1038/s41586-020-2828-1}

\bibitem[{Zhang {et~al.}(2020)Zhang, Yi, \& Wang}]{Zhang:2020cfp}
Zhang, G.~Q., Yi, S.~X., \& Wang, F.~Y. 2020, Astrophys. J., 893, 44,
  \dodoi{10.3847/1538-4357/ab7c5c}

\bibitem[{Zhu {et~al.}(2018)Zhu, Thrane, Os\l{}owski, Levin, \&
  Lasky}]{PhysRevD.98.043002}
Zhu, X., Thrane, E., Os\l{}owski, S., Levin, Y., \& Lasky, P.~D. 2018, Phys.
  Rev. D, 98, 043002, \dodoi{10.1103/PhysRevD.98.043002}

\end{thebibliography}
\end{CJK*}
\end{document}